# A Control-Oriented Framework for Coupling Physics-Based and Data-Driven Models

**Leeroy Makusha*, Preston Abadie*, Donald J. Docimo***

**Department of Mechanical & Aerospace Engineering, Texas Tech University, Lubbock, TX 79407, USA*
*(email: Leeroy.Makusha@ttu.edu, Preston.Abadie@ttu.edu, Donald.Docimo@ttu.edu)*

**Abstract**: This paper explores the coupling of dissimilar dynamic models for system-level analysis. Design, control, and estimation for dynamic systems require accurate and analytically tractable representations. However, development of such representations is impeded by the heterogeneous modeling paradigms used for subsystems and components. Prior efforts to address this provide frameworks to link model forms together, though these can neglect integration of data-driven models or relinquish features that support analysis. This work addresses this by introducing a control-oriented framework to couple two types of representations: physics-based models, and a class of data-driven models. The framework consists of a multistep procedure that transforms subsystem models into matching representations, constructs a fully coupled model using defined coupling terms, and analyzes control properties. A physics-based microgrid with a data-driven data center load model is used to demonstrate the proposed methodology and the foundation it provides to preserve analytical tractability. Results from the case study show that the coupling structure influences the equilibrium points and stability properties of the integrated system.

## 1. INTRODUCTION

Predictive models play a key role in control, estimation, and design processes for dynamic systems. While specifications vary by application, a common requirement is that the selected model paradigm facilitates analysis – the ubiquity of state space representations arguably comes from the inherent ability to quantify stability, controllability, and observability. A second common requirement is that predictions produced by the model are sufficiently accurate. However, meeting these requirements is progressively challenging. Consider microgrids, electric vehicles, and data centers, all with increasingly complex physical designs and controllers implemented in pursuit of higher measures of efficiency. The addition of new components to these systems leads to *(i)* nonlinear, uncertain, or poorly understood behavior and *(ii)* subsystems captured through different modeling paradigms. To address these challenges and continue to meet the two common requirements, research efforts have explored *data-driven modeling* and *heterogeneous model coupling*.

Data-driven modeling methods are widely used to address the first challenge and improve predictive accuracy when mathematical representation of behavior is incomplete (Zhou *et al.*, 2022). Black-box models are appropriate when there is no physical insight. Such data-reliant approaches have been applied successfully to various fields, including finance, healthcare, computer vision, and control engineering (Ljung, 2001; Hassija *et al.*, 2024). Neural network black-box approaches include recurrent architectures (Cho *et al.*, 2014), formulations that learn corrections to existing representations (He *et al.*, 2016), and neural ordinary differential equations (ODEs) that parametrize state derivatives with neural networks (Al-Janahi, Ko and Zhu, 2026). Grey-box models are an option when some physical insight is available, but several parameters remain to be determined from observed data. Techniques for this have been refined, from early neural network-augmented first-principles models (Psichogios and Ungar, 1992) to modern physics informed neural networks (Mousavi *et al.*, 2025) and universal differential equations (El-Gazzar and van Gerven, 2025).

Heterogeneous model coupling techniques provide structure to address the second challenge and connect subsystem models describing different domains. Coupling provides a more complete picture of how the dynamic system behaves. Multiphysics modeling encapsulates a broad number of methods to interconnect models, often with coupled representations having a differential algebraic equation (DAE) structure (Samin *et al.*, 2007; Kolmbauer *et al.*, 2022). Bond graphs and graph-based models represent multi-domain systems through energy-based interconnection structures that describe the flow and transformation of power between system components (Gawthrop and Bevan, 2007; Ouedraogo and Docimo, 2025). Port-Hamiltonian formulations generalize energy-based models by representing multi-domain systems with a framework that preserves system structure, power continuity, and passivity (Van Der Schaft and Jeltsema, 2014).

While significant progress has been made with respect to data-driven modeling and heterogeneous model coupling, these areas have largely developed in parallel. A limited number of studies have started to bridge the gap, including efforts toward linear system coupling (Xie, Zhang and Ilić, 2012). This paper further unifies these areas by centering its contributions on coupling nonlinear, heterogeneous models, including physics-based representations and a class of data-driven artificial neural networks (ANNs). To this end, a step-by-step framework is developed to transform the models, connect them, and, most critically, facilitate control-oriented analysis of the coupled dynamic systems. The proposed framework is applied to a microgrid powering a data center, which supports identification of key insights regarding how coupling terms influence dynamic behavior.

The remainder of this paper is organized as follows. In Section 2, an overview of the motivating energy system problem is provided, alongside data-driven and physics-based subsystem models. Section 3 presents the model coupling framework and applies it to the example system. Section 4 provides analysis of the coupled data-driven and physics-based system model. In Section 5, insights drawn from the study are summarized.

## 2. PROBLEM STATEMENT

To elucidate on the need for extensions to model coupling methods, an example energy system problem is defined. Consider a control engineer tasked with maintaining the stable operation of a microgrid using a model-based controller. The engineer's company owns the energy generation, storage, and conversion technology, and thus has provided a parameterized equivalent circuit model (ECM). However, the primary connected load is a data center owned by the company's customer. A physics-based model of the load is not available, but the engineer's colleague fit a data-driven ANN using available input and output load measurements.

This creates a modeling challenge: the microgrid dynamics are amenable to control-oriented analysis, but the data-driven load model is not readily compatible. Coupling these two models becomes nontrivial. The objective of this work is to develop a framework that enables engineers in similar situations to this example problem. The framework should couple models of different forms while preserving or furnishing structural properties required for controller design and analysis.

### *2.1 Physics-Based Microgrid Model*

Figure 1 presents the microgrid ECM. The model, drawn from (Jahan, Ouedraogo and Docimo, 2024), contains a battery pack, photovoltaic (PV) subsystem, power converters, and a voltage bus. Groups of $N_p = 400$ battery cells are connected in parallel to make a module, and $N_s = 100$ modules are connected in series to make the battery pack. Each cell has a capacity $Q = 1.21$ Ah, relaxation capacitance $C_{2,bat} = 15{,}000$ F, and resistances $R_{1,bat} = 0.0198\ \Omega$ and $R_{2,bat} = 0.2331\ \Omega$. The states include the state of charge ($SOC$) and the second capacitor's charge ($q_{2,bat}$). The battery current $I_{bat}$ is a function of the open-circuit voltage curve ($V_{oc}$) presented in (Abadie and Docimo, 2022). The battery pack dynamics are:

$$\begin{aligned} \dot{SOC} &= -\frac{I_{bat}}{N_p Q}, \\ \dot{q}_{2,bat} &= -\frac{I_{bat}}{N_p} - \frac{q_{2,bat}}{R_{2,bat}C_{2,bat}}, \\ I_{bat} &= \frac{N_p}{R_1}\left(V_{OC}(SOC) + \frac{q_{2,bat}}{C_2} - \frac{V_{bat}}{N_s}\right). \end{aligned} \quad (1)$$

The direct current (DC) buck converter steps down the battery voltage $V_{bat}$ through control of the duty cycle $D_{sd}$. The converter dynamics are described by current $I_{L,sd}$ and voltages $V_{bat}$ and $V_{c2,sd}$, correlated with capacitances $C_{1,sd} = C_{2,sd} = 1{,}000$ F and inductance $L_{sd} = 1$ H. With resistances $R_{1,sd} = R_{2,sd} = 10{,}000\Omega$ and $R_{3,sd} = 0.0050\Omega$, the dynamics are:

$$\begin{aligned} \dot{I}_{L,sd} &= \frac{1}{L_{sd}}\left(D_{sd}V_{bat} - R_{3,sd}I_{L,sd} - V_{c2,sd}\right), \\ \dot{V}_{bat} &= \frac{1}{C_{1,sd}}\left(I_{bat} - D_{sd}I_{L,sd} - \frac{V_{bat}}{R_{1,sd}}\right), \\ \dot{V}_{c2,sd} &= \frac{1}{C_{2,sd}}\left(I_{L,sd} - I_{L2,bus} - \frac{V_{c2,sd}}{R_{2,sd}}\right). \end{aligned} \quad (2)$$

The PV subsystem consists of $N_{P,PV} = 100$ solar panels in parallel. With each panel producing current $I_{PV}$, the total current injected into the microgrid is $N_{P,PV}I_{PV}$. As shown in (3), the thermal dynamics of a PV panel describe how its temperature $T_{PV}$ evolves as a function of irradiation $G$, ambient temperature $T_\infty$, and PV voltage $V_{PV}$. With the nonlinear dependency of $I_{PV}$ on $V_{PV}$, $T_{PV}$, and $G$ defined in (Villalva, Gazoli and Ruppert Filho, 2009), parameters include thermal capacitance $C_{PV} = 4{,}580$ J/K, absorptivity $\alpha = 0.7$, panel surface area $A_s = 0.8\ \mathrm{m}^2$, and convection coefficient $h = 13.39\ \mathrm{W/(m^2K)}$.

$$\dot{T}_{PV} = \frac{\alpha A_s G - hA_s(T_{PV} - T_\infty) - V_{PV}I_{PV}(V_{PV}, T_{PV}, G)}{C_{PV}} \quad (3)$$

The DC boost converter steps PV voltage up through control of the duty cycle $D_{su}$. Using the same parameters as the buck converter, the dynamics are:

$$\begin{aligned} \dot{I}_{L,su} &= \frac{1}{L_{1,su}}\left(V_{PV} - R_{3,su}I_{L,su} - D_{su}V_{c2,su}\right), \\ \dot{V}_{PV} &= \frac{1}{C_{1,su}}\left(N_{PV,P}I_{PV}\left(I_{PV}, G, V_{PV}\right) - I_{L,su} - \frac{V_{PV}}{R_{1,su}}\right), \\ \dot{V}_{c2,su} &= \frac{1}{C_{2,su}}\left(D_{su}I_{L,su} - I_{L1,bus} - \frac{V_{c2,su}}{R_{2,su}}\right). \end{aligned} \quad (4)$$

The voltage bus acts as the central hub of the microgrid. Two inductances, $L_{1,bus} = L_{2,bus} = 1$ H, are modeled to capture currents $I_{L1,bus}$ and $I_{L2,bus}$. With capacitance $C_{bus} = 100$ F and resistances $R_{1,bus} = R_{2,bus} = 0.0001\ \Omega$ for inefficiencies,

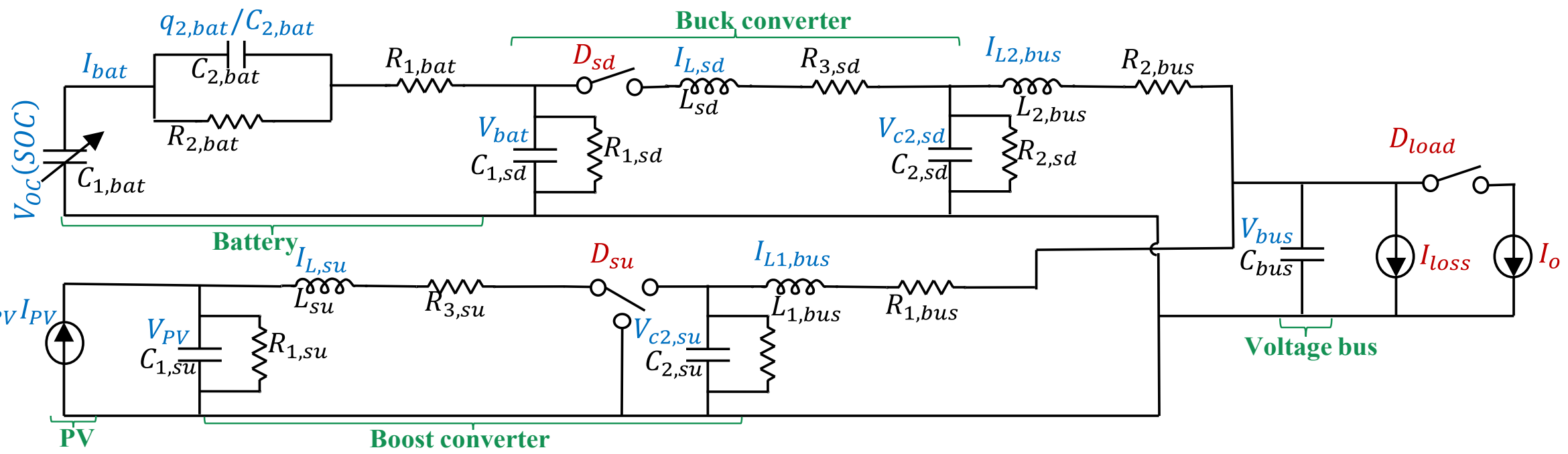

Figure 1. Microgrid equivalent circuit, with blue for voltages and currents and red for control and exogeneous inputs.

(5) describes the bus dynamics. Equation (5) contains two additional terms that affect bus voltage: $D_{load}I_O$ and $I_{loss}$. The first of these terms is current draw $I_O$ from the microgrid to the data center. The duty cycle $D_{load}$ is controlled to regulate electrical power to the load, $D_{load}I_O V_{bus}$. The second of these terms represents unknown and potentially parasitic effects created by connecting to the load.

$$\begin{aligned} \dot{I}_{L1,bus} &= \frac{1}{L_{1,bus}}\left(V_{c2,su} - V_{bus} - R_{1,bus}I_{L1,bus}\right) \\ \dot{I}_{L2,bus} &= \frac{1}{L_{2,bus}}\left(V_{c2,sd} - V_{bus} - R_{2,bus}I_{L2,bus}\right) \\ \dot{V}_{bus} &= \frac{1}{C_{bus}}\left(I_{L1,bus} + I_{L2,bus} - I_{loss} - D_{load}I_O\right) \end{aligned} \quad (5)$$

Equation (6) presents a compact version of this ECM's mathematical representation. The state vector $x_{MG}$ includes the twelve terms with time derivatives in (1)-(5), such as $T_{PV}$ and $V_{bus}$. The control input vector is $u_{MG} = [D_{sd}, D_{su}, D_{load}]$ and the exogeneous input vector is $d_{MG} = [G, T_\infty, I_O, I_{loss}]$. The vector of functions $f_{MG}$ is obtained from (1)-(5).

$$\dot{x}_{MG} = f_{MG}(x_{MG}, u_{MG}, d_{MG}) \quad (6)$$

### *2.2 Data-Driven Load Model*

Figure 2 presents the feedforward ANN that describes the nonlinear data center load dynamics. Feedforward ANNs have been shown to represent the dynamics of nonlinear systems (Pan and Duraisamy, 2018). The input $u_{DC}$ represents cooling power into the data center in kW, and the output $y_{DC}$ is the average rack temperature within the data center. At time index $k$, the network receives $N$ delayed input and output signals, $[u_{DC,k-1}, \ldots, u_{DC,k-N}]$ and $[y_{DC,k-1}, \ldots, y_{DC,k-N}]$, which are processed through functions $[\Phi_1, \ldots, \Phi_N]$. Each neuron computes a nonlinear transformation of a weighted combination of the delayed inputs and outputs. The outputs of the hidden neurons are then linearly combined to produce the predicted output.

Existing studies use networks with paired delays $u_{DC,k-n}$ and $y_{DC,k-n}$ per neuron $n$ to capture dynamic behavior (Vassiljeva, Petlenkov and Belikov, 2010). Figure 2 presents an extended form of this, where prior output $y_{DC,k-n}$ cascades down into $\Phi_{n+1}$. This is referred to as the waterfall ANN (WANN) for short, with dynamics expressed as:

$$\begin{aligned} y_{DC,k} &= \Phi_1\left(y_{DC,k-1}, u_{DC,k-1}\right) \\ &\quad + \Phi_2\left(y_{DC,k-2}, y_{DC,k-1}, u_{DC,k-2}\right) \\ &\quad + \cdots + \Phi_N\left(y_{DC,k-N}, y_{DC,k-(N-1)}, u_{DC,k-N}\right). \end{aligned} \quad (8)$$

For the present study, $N = 3$ and functions $[\Phi_1, \ldots, \Phi_N]$ are:

$$\begin{aligned} \Phi_1(\sigma_1, \sigma_2) &= w_{1,2}\sigma_1 + w_{1,1}\sigma_2 + b_1, \\ \Phi_2(\sigma_1, \sigma_2, \sigma_3) &= w_{2,2}\sigma_1 + w_{2,3}\sigma_2 + w_{21}\sigma_3 \frac{1}{1+e^{-0.5\sigma_2+10}}, \\ \Phi_3(\sigma_1, \sigma_2, \sigma_3) &= w_{3,2}\sigma_1 + w_{3,3}\sigma_2 + w_{3,1}\sigma_3 + b_2. \end{aligned} \quad (9)$$

with parameters $w_{1,2} = 1$, $w_{2,1} = -9.7182 \times 10^{-6}$, $w_{3,2} = -1.1106 \times 10^{-6}$, $b_1 = 2.2213 \times 10^{-4}$, and $b_2 = 2.7766 \times 10^{-5}$. All other weights in (9) and Figure 2 are set to zero. The model operates using a fixed timestep $\Delta t = 0.001$ s. As outlined earlier, this work emphasizes coupling a provided, pre-trained ANN to other models. Model parameters were selected to reflect approximate trends of data center cooling behavior, permitting use of advanced machine learning techniques when training becomes the focus.

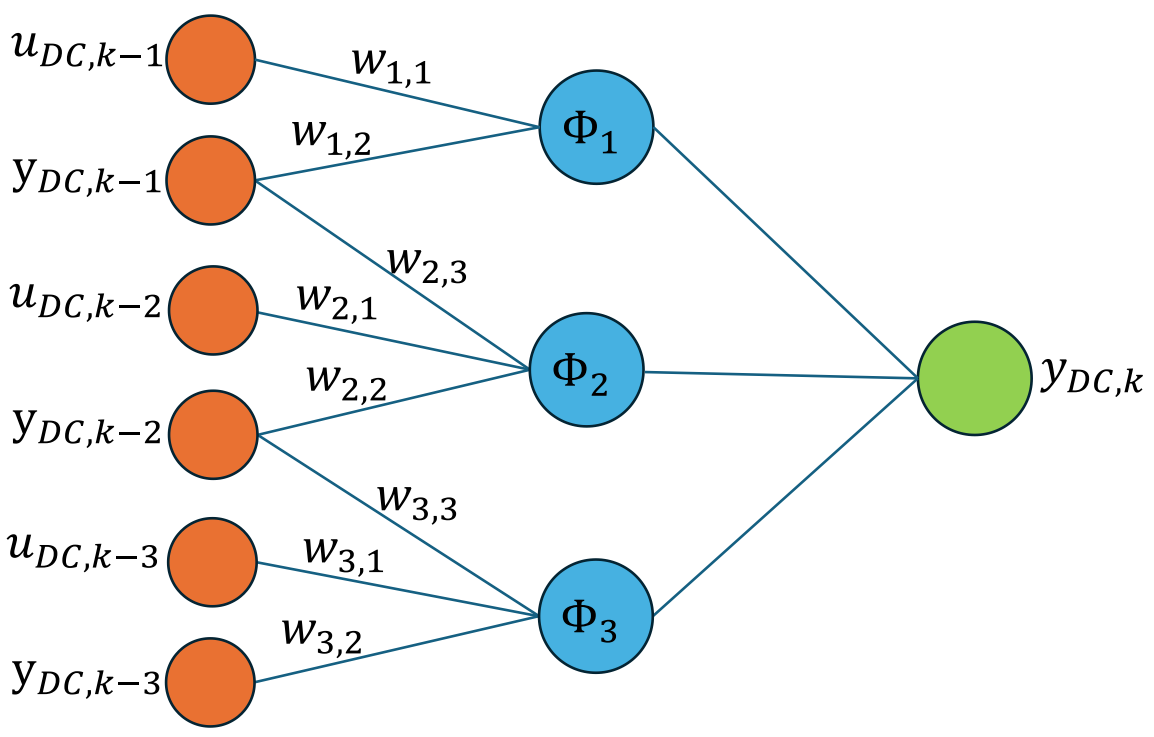

Figure 2. ANN capturing data center dynamics.

## 3. MODELING FRAMEWORK

This section presents the proposed framework to combine and analyze physics-based and WANN data-driven models. The example energy system problem is used to demonstrate the application of the framework.

### *3.1 Step 1: Transform Subsystem Models into Matching Representations*

The framework begins by transforming all individual subsystem models into the same modeling paradigm. The discrete-time state space representation is selected as the target model form due to two reasons: the framework is aimed at facilitating control-oriented analysis, and data-driven dynamic models are often obtained in discrete time.

There are a variety of established techniques for extracting state space forms of physics-based models. The model of the microgrid in Section 2.1 provides one example, converting the circuit into the state space model of (6) using Kirchoff's voltage and current laws. Linear and nonlinear partial differential equations can be converted using the Padé approximation and the finite difference method, respectively (Rahn and Wang, 2013), though care should be taken due to the error these can introduce. Once a continuous state space model is obtained, the forward Euler method or equivalent can be used to convert to discrete time. Ideally, the timestep $\Delta t$ should be selected to be consistent with other subsystem models. For the microgrid model, (6) is converted into the following, with vector of functions $F_{MG}$:

$$\begin{aligned} x_{MG,k} &= x_{MG,k-1} + \Delta t f_{MG}\left(x_{MG,k-1}, u_{MG,k-1}, d_{MG,k-1}\right) \\ &= F_{MG}\left(x_{MG,k-1}, u_{MG,k-1}, d_{MG,k-1}\right). \end{aligned} \quad (10)$$

The conversion of data-driven models is not always as straightforward. While neural ODEs can be discretized (Chen *et al.*, 2018) and some representations are fitted effectively as discrete-time state space models with hidden states (Karl *et al.*, 2017), many practical data-driven models have non-obvious state space representations. Consider the transformation of the single input, single output WANN of (8). The selection of a set of state variables, $x_{DC} = [x_{DC,1}, \ldots, x_{DC,N}]$, to yield a minimal realization is not obvious due to the nonlinearities and multiple input delays. As discussed in (Kotta and Sadegh, 2002), a state space realization is not guaranteed for all but a subset of nonlinear dynamic models. Following that paper's work, the WANN can be shown to fit within a subclass of models that is transformable. The discrete-time state space model for (8) is:

$$\begin{aligned} x_{DC,1,k} &= x_{DC,2,k-1} + \Phi_{1,k-1}, \\ x_{DC,2,k} &= x_{DC,3,k-1} \\ &\quad + \Phi_2\big(x_{DC,1,k-1}, x_{DC,2,k-1} + \Phi_{1,k-1}, u_{DC,k-1}\big), \\ &\;\;\vdots \\ x_{DC,N-1,k} &= x_{DC,N,k-1} \\ &\quad + \Phi_{N-1}(x_{DC,1,k-1}, x_{DC,2,k-1} + \Phi_{1,k-1}, u_{DC,k-1}), \\ x_{DC,N,k} &= \Phi_N\big(x_{DC,1,k-1}, x_{DC,2,k-1} + \Phi_{1,k-1}, u_{DC,k-1}\big) \end{aligned} \tag{11}$$

with $y_{DC} = x_{DC,1}$ and $\Phi_{1,k-1} = \Phi_1\big(x_{DC,1,k-1}, u_{DC,k-1}\big)$. Here $x_{DC,1}$ represents the rack temperature and $\big[x_{DC,2}, \dots, x_{DC,N}\big]$ are latent states to capture the underlying nonlinear dynamics. Using $F_{DC}$, this can be shortened to:

$$x_{DC,k} = F_{DC}\big(x_{DC,k-1}, u_{DC,k-1}\big). \tag{12}$$

### *3.2 Step 2: Define Coupling Terms*

The purpose of this step is to establish representations of interactions among the subsystem models produced through the first step. Coupling arises from shared inputs, outputs, and internal variables. For this preliminary version of the framework, four guidelines are recommended:

1. Each set of coupling terms only involves variables from two models.
2. Each coupling term is an equation with a single input variable on the left side. The equation defines how a subsystem "receives" effects from another subsystem.
3. No input variable acts as a receiver more than once.
4. No input variable is a function of itself, whether directly or indirectly through the set of coupling equations.

To show how these guidelines work, the set of coupling terms for the microgrid and data center models are defined. The cooling power input to the data center, $u_{DC}$, is related to the electrical power supplied through the microgrid's voltage bus. Equation (13) describes the first coupling term, from microgrid to data center, with a constant coefficient of performance ($COP$). Similarly, the data center load influences the microgrid through $I_O$. Equation (14) presents the second coupling term, from data center to microgrid, with $R_{DC}$ as the effective load resistance of the data center cooling system when it draws power. Note that these coupling terms meet all four guidelines. If (14) were to violate the last guideline and have $I_O$ as a function of $u_{DC}$, the created algebraic relationship should be solved before proceeding.

$$u_{DC} = COP \cdot D_{load} V_{bus} I_O \tag{13}$$

$$I_O = V_{bus}/R_{DC} \tag{14}$$

One more coupling term is necessary to connect the microgrid and data center models. The exogeneous input $I_{loss}$ captures additional effects of the data center on the microgrid. This can include current drawn to power auxiliary loads and parasitic terms. As these can be influenced by the thermal state of the data center, the coupling term is defined as:

$$I_{loss} = V_{bus} H\big(x_{DC,1}\big), \tag{15}$$

with $H$ as the inverse of a temperature-dependent impedance. This is not a physical resistor, but rather an aggregated interaction capturing both passive and active electrical effects.

### *3.3 Step 3: Construct the Fully Coupled Model*

After the subsystem models have been transformed and the coupling terms defined, what follows is construction of coupled model describing the full system. The new definitions for input variables, provided by the coupling term equations, are placed within the subsystem models. The subsystem state vectors are combined into one vector, $x$. Vectors $u$ and $d$ are similarly defined, excluding inputs replaced through use of the coupling equations. For the microgrid and data center case, the coupled state equations can be written as:

$$\begin{aligned} x_{MG,k} &= F_{MG}\big(x_{MG,k-1}, u_{MG,k-1}, d_{MG,k-1}\big), \\ x_{DC,k} &= F_{DC}\big(x_{DC,k-1}, COP \cdot D_{load} V_{bus} I_O\big), \end{aligned} \tag{16}$$

with $d_{MG,k-1} = \Big[G_{k-1}, T_{\infty,k-1}, \frac{V_{bus,k-1}}{R_{DC}}, V_{bus,k=1} H\big(x_{DC,k-1}\big)\Big]$. This can be summarized as:

$$x_k = F(x_{k-1}, u_{k-1}, d_{k-1}), \tag{17}$$

with $x = [x_{MG}, x_{DC}]$, $u = u_{MG}$, and $d = [G, T_\infty]$.

### *3.4 Step 4: Analyze Control Properties*

A coupled model in the form of (17) provides access to all tools appropriate for nonlinear, discrete-time state space models. The final step of the framework is to apply these tools as appropriate for the problem at hand. Table 1 presents a family of suggested metrics to serve as a starting point for control-oriented analysis of the coupled system. Methods are not limited to Table 1: for example, input-to-state stability analysis can also be employed (Jiang and Wang, 2000).

## 4. RESULTS & DISCUSSION

This section covers the control-oriented analysis of the coupled microgrid and data center system model. The first two analyses of Table 1 are performed: equilibrium is determined, and local stability is quantified. For this section, coupling parameters are set to $COP = 3.5$ and $R_{DC} = 3.7\ \Omega$. Two expressions for $H$ are utilized, with the equation and curves presented in Figure 3. Coupling Case A uses $\gamma = 0.005$ and Coupling Case B uses $\gamma = -1$. This latter case captures a scenario where auxiliary loads feed current back into the bus at higher rack temperatures. This behavior is analogous to negative incremental impedance effects that have been observed in power electronics, where load dynamics can destabilize upstream converters (Emadi *et al.*, 2006).

Table 1. Suggested analyses to perform using the coupled model

| Analysis Name | Method |
|---|---|
| Equilibrium | Set $x_k = x_{k-1}$, select free variables, and solve using analytical & numerical methods. Compare uncoupled vs. coupled model equilibrium points. |
| Stability (Local) | Determine the Jacobian at equilibrium & calculate eigenvalues. Perturb coupling terms and observe influence on eigenvalues. |
| Stability (Global) | Construct a radially unbounded Lyapunov function $V(x)$ around equilibrium $\bar{x}$ such that $V(x) > 0$ for $x \neq \bar{x}$ and $V\big(F(x)\big) - V(x) < 0$ for all $x$. |
| Controllability | Linearize around equilibrium $(\bar{x}, \bar{u})$ by computing: $A = \frac{\partial F}{\partial x}, B = \frac{\partial F}{\partial u}$, and testing rank of controllability matrix: $[A^0 B \quad \cdots \quad A^{n-1} B]$ |

### *4.1 Equilibrium Analysis*

An equilibrium point is determined by setting $x_k = x_{k-1} = \bar{x}$, $u_{k-1} = \bar{u}$, and $d_{k-1} = \bar{d}$. There are five free variables in the coupled model, which are set to $\bar{u} = [0.5, 0.15, 1.0]$ and $\bar{d} = [1000, 25]$. Using a combination of analytical and numerical methods, (17) is solved for these conditions for both coupling cases. In addition, an equilibrium point is also determined for the uncoupled subsystem models: this requires selection of additional free variables $\bar{I}_O = 20.5$ A, $\bar{I}_{loss} = 0.0165$ A, and $\bar{u}_{DC} = 31$ kW. Table 2 presents values of key states at an equilibrium point identified.

Two observations of note are extracted from this table. First, coupling the models is necessary to get an accurate view of equilibrium, shown by comparing the coupling and uncoupled cases. If the control engineer in this example scenario were to use only the microgrid model, it would be necessary to select $\bar{I}_O$ and $\bar{I}_{loss}$. Knowledge of proper values of these comes from the coupling – without it, the engineer is unlikely to select values reflective of realistic conditions. This is why the values are significantly different for coupled vs. uncoupled. Second, the form of the coupling terms has a substantial effect on equilibrium, shown by comparing the two coupling cases in Table 2. The change in $\gamma$ shifts the equilibrium and even provides an indicator of the limits of the original subsystem model's accuracy. For example, consider the negative $V_{bus}$ equilibrium value for Case B. This is impractical for a real-world scenario, indicating that the physical system could break down at this point due to unmodeled nonlinearities. This gives reason to explore the stability of the coupled model.

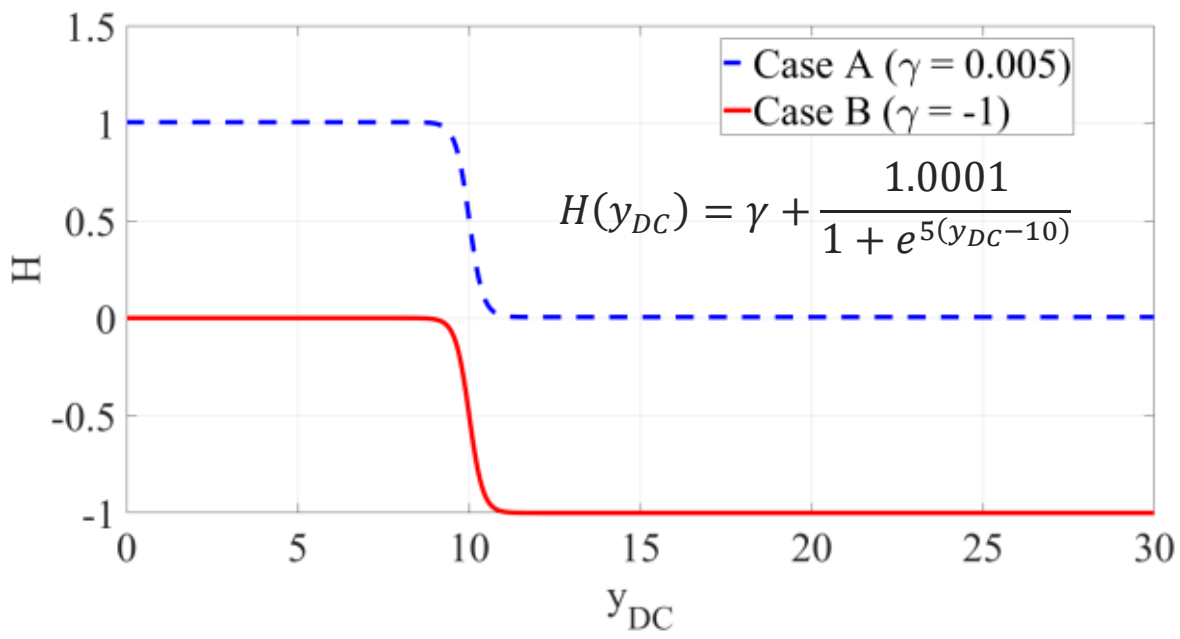

Figure 3. Coupling term $H$ for the two cases analyzed, A and B.

Table 2. Value of key states at equilibrium

| States | Coupled A ($\gamma = 0.005$) | Coupled B ($\gamma = -1$) | Uncoupled |
|---|---|---|---|
| $SOC$ [−] | 0.1876 | -0.0769 | 0.9663 |
| $V_{bat}$ [V] | 232.5424 | -0.4114 | 339.9540 |
| $V_{PV}$ [V] | 25.7540 | -0.0259 | 26.1842 |
| $V_{bus}$ [V] | 161.7716 | -0.2057 | 169.9773 |
| $x_{DC,1}$ [°C] | 23.1065 | 112.5042 | 21.4308 |

### *4.2 Stability Analysis*

Figure 4 presents a simulation of the models for Coupling Cases A and B, with constant control and exogeneous inputs at $\bar{u}$ and $\bar{d}$. The variation of the coupling term $\gamma$ within $H$ seems to affect stability of the system. To quantify stability, the eigenvalues of the Jacobian of (17), $\partial F / \partial x$ evaluated at $[\bar{x}, \bar{u}, \bar{d}]$, are determined. This is an appropriate measure of stability for a nonlinear system about equilibrium (Levine, 2011) – for discrete-time systems, eigenvalues need to be within the unit circle for stability.

Figure 5 presents the eigenvalues for the earlier equilibrium with respect to both coupling cases. From the zoomed-in portion, all eigenvalues for Coupling Case A are within the unit circle, indicating stability. However, three eigenvalues are outside the unit circle for Coupling Case B, syncing with the unstable trajectory observed in Figure 4. The magnitude is approximately 1.000003. This could indicate a few things, including numerical error from discretization. However, it also can be indicative of slow operating timescales, which is feasible here due to the large energy storage terms. This preliminary analysis can be complemented using additional tools for stability analysis as mentioned earlier. A trend can be extracted, shown in Figure 5's zoomed-in portion, relating the eigenvalues as a function of $\gamma$. This opens the path to a deeper sensitivity analysis if the engineer requires it.

## 5. CONCLUSIONS

This work introduces and applies a control-oriented framework to couple physics-based models and a class of data-driven models. This framework enables unified modeling and systematic analysis of key control properties in heterogeneous dynamic systems. Its effectiveness is demonstrated through the integration of a physics-based microgrid model and a data-driven data center load model. Equilibrium and stability analyses indicate that the coupling can significantly shift the equilibrium points and, in some cases, destabilize the overall system. The framework provides critical insights for controller design by revealing system-level interactions that would remain hidden without proper coupling.

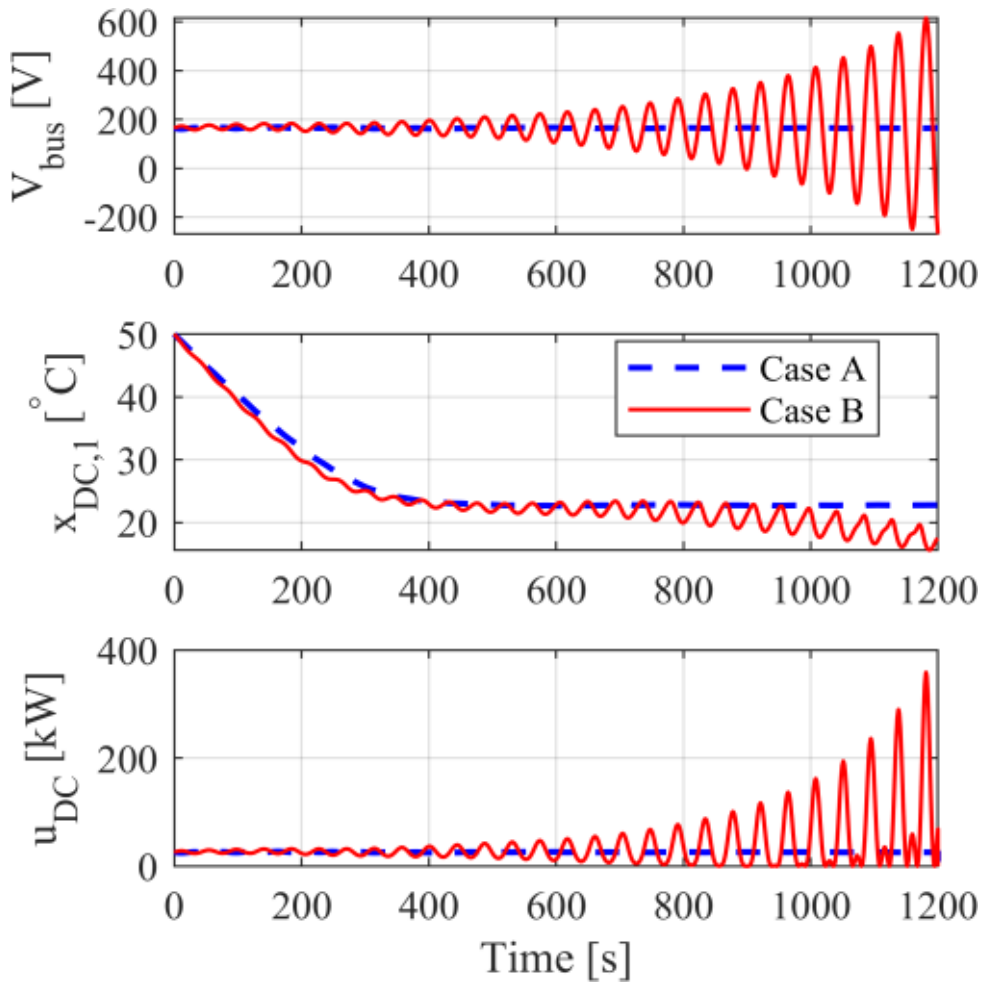

Figure 4. Simulation results of the coupled microgrid and data center model for case A and case B.

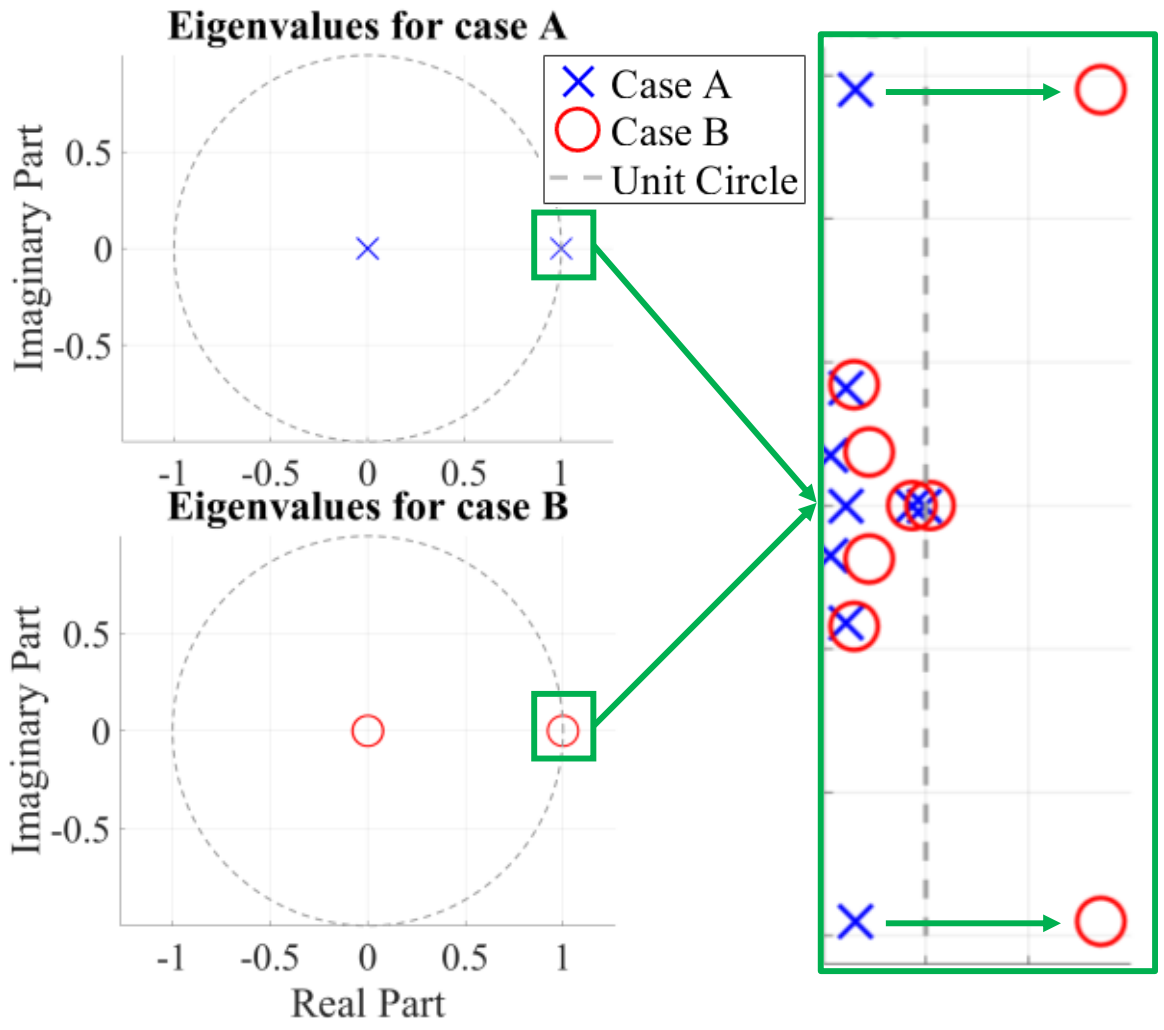

Figure 5. Eigenvalue stability tests for case A and case B.

ACKNOWLEDGEMENTS

This material is based upon work supported by the National Science Foundation under Award No. 2324707.